# Modeling for Semiconductor Spintronics

Semion Saikin[1,2], Yuriy V. Pershin[1,3] and Vladimir Privman[1]

[1] *Center for Quantum Device Technology,*
*Clarkson University, Potsdam, NY 13699-5721, USA*
[2] *Department of Theoretical Physics,*
*Kazan State University, Kazan 420008, Russia*
[3] *Department of Physics and Astronomy, Michigan State University,*
*East Lansing, Michigan 48824-2320, USA*

We summarize semiclassical modeling methods, including drift-diffusion, kinetic transport equation and Monte Carlo simulation approaches, utilized in studies of spin dynamics and transport in semiconductor structures. As a review of the work by our group, several examples of applications of these modeling techniques are presented.

**I. Introduction**

The idea to use the spin property of electrons in semiconductor electronic devices (semiconductor spintronics) is considered as one of the promising trends for future electronics [1,2]. It is expected that utilization of spin-related phenomena in information processing will extend the functionality of conventional devices and allow development of novel electronic devices based on new operating principles [3-9]. Recent interest has been motivated by successful examples of metallic spintronic devices, such as ferromagnetic metal-based reading heads for hard disc drives and magnetic random access memory. Possible applications of semiconductor spintronics include magnetic sensing and non-volatile magnetic memory. In comparison with metal-based spintronics, recently reviewed, e.g., in [10], utilization of semiconductor structures promises more versatile design due to the ability to adjust potential variation and spin polarization in the device channel by external voltages, device structure and doping profiles. Different types of devices [11-25] have been proposed recently. However, the actual advantages of most of these designs as compared to the conventional electronics devices have not been clearly established [26]. For example, the most promising proposal for an electronic analogue of electro-optical modulator [13] that was later termed "spin-field effect transistor," has been criticised in [27]. To resolve this controversy a substantial basic research effort is required. Device design plays an important role in this problem [24,25].

Recent experimental advances have allowed generation and control of non-equilibrium spin polarization in semiconductor structures [4-9]. For example, it has been demonstrated that spin polarization can be maintained [28] for up to several nanoseconds at room temperature and efficiently controlled by gate voltage [29] in GaAs (110) quantum wells. Coherent injection of polarized spins across material interfaces [30] and coherent transport of spin polarization in homogeneous materials for a distance longer than 100 micrometers [31] have been reported. Some of the experiments go beyond the application of conventional electronic devices. It has been shown that spin-polarized current can reduce the threshold of a semiconductor laser up to 50% [32]. Information about electron spin polarization can be transmitted to photon polarization [19] and decoded by a detecting device [33]. More sophisticated experiments can control pure spin currents without charge currents in bulk semiconductors [34].

The goal of this paper is to review spin transport modelling approaches in the semiclassical, high temperature regime of spatial transport, appropriate for operation of conventional devices of interest in electronics. In the following section we discuss interactions affecting spin dynamics in semiconductors. In Section III, the drift-diffusion approach to spin transport modelling is considered. We describe both the two-spin state and the spin-polarization-vector approaches, the latter offering a fuller quantum-mechanical treatment of the spin dynamics. In Section IV, a Boltzmann-like kinetic transport equation including spin dependent terms is derived based on the Wigner function approach. Section V is devoted to Monte Carlo modelling of spin transport in semiconductor structures. Throughout the text we provided several examples, based on the work by our group, illustrating the considered approaches. In most of these studies we concentrate on details of spin transport in different structures but do not try to compare spintronic and conventional devices. The latter issue requires more specific analyses of particular device designs.

**II. Spin dynamics of charge carriers**

Spin, **s**, is a quantum mechanical property, associated with intrinsic angular momentum vector of electrons, and many other elementary particles, as well as nuclei and atoms. An intrinsic magnetic moment is associated with spin. Therefore, spin is closely related to magnetic phenomena. However, the quantum mechanical nature of the spin makes it different from the classical angular momentum. The most important is that the spin is strongly quantized and can only take on discrete values. Its vector components do not commute with each other. Therefore, electron spin requires in most cases a fully quantum-mechanical treatment. A natural constant which arises in the treatment of magnetic properties of electrons—spin ½ particles—is the Bohr magneton, $\mu_B = 9.274 \cdot 10^{-24}$ J/T.

In semiconductor spintronic structures, where spin is carried by electrons and/or holes, the spin dynamics is controlled by magnetic interactions. Some of these are surveyed below.

*Interaction with an external magnetic field.*

An external magnetic field $\vec{B}$ exerts a torque on a magnetic dipole and the magnetic potential energy is given by Zeeman term

$$U = \frac{g^* \mu_B}{2} \vec{\sigma} \cdot \vec{B} \, , \qquad (1)$$

where $g^*$ is the effective $g$-factor, and $\vec{\sigma}$ represents a vector of the Pauli spin matrices, used in the quantum-mechanical treatment of spin ½, see [35]. The interaction (1) leads to the spin precession around the external magnetic field. This interaction is important in all system where a magnetic field is present. Moreover, fluctuations of $\vec{B}$ could lead to noise contributing to spin relaxation.

*Interaction with magnetic impurities, nuclei and other spin carriers.*

An electron located in a semiconductor experiences different kinds of spin-spin interactions including direct dipole-dipole interactions with nuclear spins and other (free and localized) electrons, and the exchange interaction. The latter, in fact, is the result of the electrostatic Coulomb interaction between electrons, which becomes spin-dependent because of the Pauli exclusion principle [35]. Usually, at room temperatures in sufficiently clean, low-doped non-magnetic semiconductors these interactions are not very important.

*Spin-orbit interaction.*

The spin-orbit (SO) interaction arises as a result of the magnetic moment of the spin coupling to its orbital degree of freedom. It is actually a relativistic effect, which was first found in the emission spectra of hydrogen. An electron moving in an electric field, sees, in its rest frame, an effective magnetic field. This field, which is dependent on the orbital motion of the electron, interacts with the electron's magnetic moment.

The Hamiltonian describing SO interaction, derived from the four-component Dirac equation [36], has the form

$$H_{SO} = \frac{\hbar^2}{4m^2c^2} \left( \vec{\nabla} V \times \vec{p} \right) \cdot \vec{\sigma} \, , \qquad (2)$$

where $m$ is the free electron mass, $\vec{p}$ is the momentum operator, and $\vec{\nabla} V$ is the gradient of the potential energy, proportional to the electric field acting on the electron. When dealing with crystal structures, the spin orbit interaction, Eq. (2), accounts for symmetry properties of materials. Here we emphasise two specific mechanisms that are considered to be important for spintronics applications. The Dresselhaus spin-orbit interaction [37] appears as a result of the asymmetry present in certain crystal lattices, e.g., the zinc blende structures. For a two-dimensional electron gas in semiconductor heterostructures with an appropriate growth geometry, the Dresselhaus SO interaction is of the form

$$H_D = \frac{\beta}{\hbar}\left(\sigma_x p_x - \sigma_y p_y\right).\qquad(3)$$

Here, $\beta$ is the coupling constant.

The Rashba spin-orbit interaction [38] arises due to the asymmetry associated with the confinement potential and is of interest because of the ability to electrically control the strength of this interaction. The latter is utilized, for instance, in the Datta-Das spin transistor [13]. The Hamiltonian for the Rashba interaction is written [38] as

$$H_R = \frac{\alpha}{\hbar}\left(\sigma_x p_y - \sigma_y p_x\right),\qquad(4)$$

where $\alpha$ is the coupling constant.

Other possible sources of spin-orbit interaction are non-magnetic impurities, phonons [39], sample inhomogeneity, surfaces and interfaces. In some situations these could play a role in spin transport and spin relaxation dynamics.

In general, transport in semiconductor spintronic devices can be characterized by the creation of a non-equilibrium spin polarization in the device (spin injection), measurement of the final spin state (spin detection), external control of the spin dynamics by the electric (gate modulation) or magnetic fields, and uncontrolled spin dynamics leading to loss of information in the device (spin relaxation or spin dissipation). Specifically, once injected into a semiconductor, electrons experience spin-dependent interactions with the environment, which cause spin relaxation. The spin polarization in a typical semiconductor at room temperature is then lost over distances no larger than a fraction of $0.1\,\mu m$. This has been the main reason that the development of semiconductor spintronics is closely linked with the advent of nanotechnology, and spin-dependent phenomena in transport will play any role only when device components reach truly deep-sub-micron dimensions.

**III. Drift-diffusion approach**

The drift-diffusion approximation is the simplest approach to spin involving process modeling. Based on the well-known drift-diffusion model used to describe phenomena related only to the charge degree of freedom, drift diffusion models accounting for the spin degree of freedom can be subdivided into two classes: two-component drift-diffusion approximations and spin-polarization-vector or density-matrix based models. Both approaches have been used successfully in practical modeling of spin-related phenomena in semiconductors [40-52]. General conditions for the applicability of these approximations are not different from the usual conditions of applicability of drift-diffusion approximations.

*Two-component drift-diffusion model.*

The two-component drift-diffusion approach, originally devised for spin transport modelling in ferromagnetic metals, has been applied to spin-related problems in semiconductors [40-46,52]. This model ignores the transverse spin coherence; spin relaxation processes are included phenomenologically. Generally, the electrons are considered to be of two types, having spin up or down. Drift-diffusion equations for each type of electron, including the relaxation terms, can be formulated as follows,

$$e\frac{\partial n_{\uparrow(\downarrow)}}{\partial t} = div\, \vec{j}_{\uparrow(\downarrow)} + \frac{e}{2\tau_{sf}}\left(n_{\downarrow(\uparrow)} - n_{\uparrow(\downarrow)}\right) + S_{\uparrow(\downarrow)}(\vec{r},t) \ , \qquad (5)$$

$$\vec{j}_{\uparrow(\downarrow)} = \sigma_{\uparrow(\downarrow)}\vec{E} + eD\nabla n_{\uparrow(\downarrow)} \ , \qquad (6)$$

and

$$\sigma_{\uparrow(\downarrow)} = en_{\uparrow(\downarrow)}\mu \ , \qquad (7)$$

where $e$ is the electron charge, $n_{\uparrow(\downarrow)}$ is the density of the spin-up (spin-down) electrons, $\vec{j}_{\uparrow(\downarrow)}$ is their current density, $\tau_{sf}$ is the spin relaxation time, $S_{\uparrow(\downarrow)}(\vec{r},t)$ describes the source of the spin polarization, $\sigma_{\uparrow(\downarrow)}$ is the conductivity, and $\mu$ is the mobility, connected with the diffusion coefficient $D$ via the Einstein relation $\mu = De/(k_B T)$, and defined via $\vec{v}_{drift} = \mu\vec{E}$.

Equation (5) is the usual continuity relation that takes into account spin relaxation and sources of the spin polarization, Eq. (6) is the expression for the current which includes the drift and diffusive terms, and Eq. (7) is the expression for the conductivity. It is assumed that the diffusion coefficient $D$ and the spin relaxation time $\tau_{sf}$ are equal for the spin-up and spin-down electrons. Eqs. (5)-(7) can be supplemented by the Poisson equation accounting for, e.g., the effects of a non-homogeneous electric field,

$$div\,\vec{E} = \frac{e}{\varepsilon\varepsilon_0}(N - n) \ . \qquad (8)$$

Here $N$ is the positive background charge density and $n = n_{\downarrow} + n_{\uparrow}$ is the electron density. From Eqs. (5)-(8), we obtain an equation for the spin polarization density $P = n_{\uparrow} - n_{\downarrow}$,

$$\frac{\partial P}{\partial t} = D\Delta P + D\frac{e\vec{E}}{k_B T}\nabla P + D\frac{e\nabla\vec{E}}{k_B T}P - \frac{P}{\tau_{sf}} + F(\vec{r},t) \ , \qquad (9)$$

which can be supplemented by an equation for the electric field [43]. Here $F(\vec{r},t) = \left[ S_\uparrow(\vec{r},t) - S_\downarrow(\vec{r},t) \right]/e$ represents a spin polarization density being created by the external source.

*Accumulation of electron spin polarization at semiconductor interfaces.*

Let us consider an example of using the two-component drift-diffusion model to describe propagation of electron spin polarization through an interface separating two n-type semiconductor regions in an applied electric field [46]. We assume that an inhomogeneous spin polarization is created locally by a continuous source and is driven through a boundary by an electric field, see Fig. 1a. Each semiconductor region is characterized by the following parameters: the diffusion coefficient $D_i$, doping level $N_i$, and spin relaxation time $\tau_i$. Neglecting the effect of charge redistribution at the interface (this effect was studied in [43]), the evolution of the electron spin polarization is described in each semiconductor by Eq. (9). These equations were solved analytically with appropriate boundary conditions, representing flux conservation and continuity of spin polarization density at the interface [46].

It was found that for certain values of parameters describing the system, the electron spin polarization is accumulated near the interface. An example of a such situation is shown in Fig. 1b, where accumulation of the electron spin polarization at the interface is more pronounced with increase of the doping density $N_2$. Qualitatively, this effect occurs because of the lower drift velocity in the higher doped region resulting in deceleration of spin-polarized electrons after passing the interface and, correspondingly, in their accumulation. In order to avoid confusion, it should be emphasized that this method allows increasing the electron spin polarization density $n_\uparrow - n_\downarrow$, but not the level of the electron spin polarization, defined as $(n_\uparrow - n_\downarrow)/(n_\uparrow + n_\downarrow)$. Moreover, we wish to note here that the spin polarization density is continuous *at* the interface; its sharp decrease in the left region near the interface is described by a fast decaying function [46]. It was shown in Ref. [46] that in order to obtain a high level of the spin polarization density at the interface, it is necessary that both semiconductors have long electron spin relaxation times; the right semiconductor should have small electron diffusion coefficient and high doping level.

*Gate control of spin polarization drag in semiconductor heterostructures.*

In this subsection, we consider an example of modelling of a gate operation in a non-ballistic spin-FET [15]. The proposed device consists of a gated heterostructure with ferromagnetic source and drain contacts. The device utilizes spin relaxation of conduction electrons in a semiconductor quantum well (QW) modulated by the gate voltage [15]. Spin-polarized electrons, injected from the source, propagate in the plane of the QW, formed in the heterostructure, and are filtered by the drain contact. The spin dynamics of the conduction electrons is controlled by the spin-orbit interaction. It was shown theoretically that for some particular configurations, spin-polarized electrons can be transported without substantial loss of polarization if the spin-orbit coupling coefficients, $\alpha$ and $\beta$, are nearly equal [15,47,53]. However, if the spin-orbit constants are made not equal, for example, by applying an external electric field orthogonal to the QW plane, the spin dephasing mechanism [54] is efficient. As a result, a different spin polarization of electrons in the device channel near the magnetic drain contact is produced. The magnetoresistance of the structure, and therefore the current through the device, is dependent on the value of this spin polarization and its relative orientation with respect to the magnetization of the drain.

One can derive that spin density along the channel for this example decays exponentially [47] as

$$P(x) = P(0)e^{-x/L_s}, \qquad (10)$$

with the characteristic spin scattering length

$$L_s = \left( \frac{\mu E}{2D} + \sqrt{\left(\frac{\mu E}{2D}\right)^2 + \left(\frac{2m^*(\alpha(V_g) - \beta(V_g))}{\hbar^2}\right)^2} \right)^{-1}, \qquad (11)$$

where spin-orbit coefficients are functions of the gate voltage. These coefficients can be calculated based on the device material band structure [52]. For different values of the gate voltage the spin density in the channel varies due to the change of the electron concentration and due to the modulation of the spin dephasing. To characterize the efficiency of the latter mechanism, one should use spin polarization, normalized to the number of the electrons. Using realistic material parameters it is possible to establish that the modulation of the spin polarization in a submicron size AlGaAs/GaAs gated heterostructure at room temperature within a reasonable range of the gate voltages is of the order of 15-20% [52]. This effect can be observed experimentally. However, for a device application further improvements of the structure design are desirable.

*Spin polarization vector approach.*

As an extension of the previously discussed model this approach accounts for the transverse spin dynamics [47-51]. This is important if time scales of processes studied are comparable with a characteristic decoherence/dephasing time for quantum-mechanical superposition of spin-up and spin-down states. For example, this transport regime can be exploited for spintronic devices operating with analog logic [55]. The spin polarization density, **P**, in this case is a vector quantity. It corresponds to a single-particle spin density matrix as

$$\rho_\sigma = \frac{1}{2} \begin{pmatrix} 1 + P_z & P_x - iP_y \\ P_x + iP_y & 1 - P_z \end{pmatrix}. \qquad (12)$$

It should be noted that representation (12) is still a single-electron description. For example, it cannot describe spin entanglement of two electrons. Under the assumption that the spin degree of freedom does not affect the spatial motion, it is possible to show that the dynamics of spin polarization density is described by a vector drift-diffusion equation. For example, in the 1D case it is [47, 49]

$$\frac{\partial \mathbf{P}}{\partial t} - \hat{\mathbf{D}}\frac{\partial^2 \mathbf{P}}{\partial x^2} - \hat{\boldsymbol{\mu}}\frac{\partial \mathbf{P}}{\partial x} + \hat{\mathbf{C}}\mathbf{P} = 0 \ . \tag{13}$$

The coefficients, $\hat{\mathbf{D}}$, $\hat{\boldsymbol{\mu}}$, $\hat{\mathbf{C}}$, in Eq. (13) are 3 by 3 matrixes in the spin space. The symmetry of these coefficients is defined by the properties of specific spin-dependent interactions. In general, all three matrix coefficients cannot be diagonalized simultaneously and equations for the components of the spin polarization density vector cannot be decoupled.

*Short-time approximation.*

In many applications, it is important to find the spatial distribution of the spin polarization, $\mathbf{P}(\mathbf{r},t)$, at an arbitrary moment of time, $t$, given the initial spin polarization distribution, $\mathbf{P}(\mathbf{r},0)$. Initial dynamics of the spin polarization distribution can be found using the short-time approximation [50]. Within this approximation, $\mathbf{P}(\mathbf{r},t)$ in the two-dimensional heterostructure geometry is given by

$$\mathbf{P}(x,y,t) = \iint G(x-x', y-y', t)\mathbf{P'}_{(x,y),(x',y')}\, dx'dy' \ , \tag{14}$$

where $G(x-x', y-y', t)$ is the diffusion Green's function (solution of the diffusion equation with a point source), and $\mathbf{P'}_{(x,y),(x',y')}$ represents the contribution of the initial spin polarization density at the point $(x',y')$ to $\mathbf{P}(x,y,t)$. The structure of Eq. (14) can be easily understood. Electron spin polarization density in a space volume with coordinates $(x,y)$ at a selected moment of time $t$ is given by a sum of the spin polarization vectors of all the electrons located in this volume. The diffusion Green's function $G(x-x', y-y', t)$ gives the probability for the electrons to diffuse from the point $(x',y')$ to $(x,y)$, while $\mathbf{P'}_{(x,y),(x',y')}$ describes the spin polarization of these electrons.

For the initial spin relaxation dynamics, the main approximations used to make Eq. (14) tractable are the assumption that different spin rotations commute with each other, and that the spin precession angle $\varphi$ is proportional to the distance between $(x',y')$ and $(x,y)$. These assumptions are justified when the spin precession angle per mean free path is small and the time is short. Moreover, it is assumed that evolution of the electron spin degree of freedom is superimposed on the space motion of the charge carriers. In other words, the influence of the spin-orbit interaction on the spatial motion is neglected. If $\mathbf{a}$ is the unit vector along the precession axis, then [50]

$$\mathbf{P'}_{(x,y),(x',y')} = \mathbf{P} + \mathbf{P}_\perp \left(\cos\varphi - 1\right) + \mathbf{a} \times \mathbf{P} \sin\varphi \ , \tag{15}$$

where $\mathbf{P}_\perp = \mathbf{P} - \mathbf{a}(\mathbf{aP})$ is the component of the spin polarization perpendicular to the precession axis, $\varphi = \eta r$, while $\eta$ is the spin precession angle per unit length, $\mathbf{r} = (x - x', y - y')$, and $r = |\mathbf{r}|$. Here $\mathbf{P} = \mathbf{P}(\mathbf{r}, t = 0)$. With only the Rashba spin-orbit interaction, $\mathbf{a} = \hat{z} \times \mathbf{r} / r$, and $\hat{z}$ is the unit vector in z direction, perpendicular to QW. The definition of $\mathbf{a}$ in a more general case is given in [49]. It should be emphasized that the spin-orbit interaction is the origin of the spin polarization rotations described by Eq. (15). The short-time approximation was used [50] in the investigation of the spin relaxation dynamics near the edge of two-dimensional electron gas (2DEG).

*Anisotropy of spin transport in 2DEG.*

A specific form of the coefficients, $\hat{\mathbf{D}}$, $\hat{\boldsymbol{\mu}}$, $\hat{\mathbf{C}}$, in the drift-diffusion equation for the spin polarization vector can be obtained by different methods [47, 49]. For example, to describe spin dynamics in 2DEG, controlled by the spin-orbit interaction, Eqs. (3)-(4), one can apply the moments expansion procedure to the Wigner function equation (see the next section) [47]. In this case the coefficients, $\hat{\mathbf{D}}$, $\hat{\boldsymbol{\mu}}$, $\hat{\mathbf{C}}$, are

$$\hat{\mathbf{D}} = \begin{pmatrix} D & 0 & 0 \\ 0 & D & 0 \\ 0 & 0 & D \end{pmatrix}, \quad \hat{\boldsymbol{\mu}} = \begin{pmatrix} \mu E & 2B_{xz}D & 0 \\ -2B_{xz}D & \mu E & 0 \\ 0 & 0 & \mu E \end{pmatrix},$$

$$\hat{\mathbf{C}} = \begin{pmatrix} D(B_{xz}^2 + B_{yz}^2) & -\mu E B_{xz} & -B_{yx}B_{yz}D \\ \mu E B_{xz} & D(B_{xz}^2 + B_{yx}^2 + B_{yz}^2) & 0 \\ -B_{yx}B_{yz}D & 0 & DB_{yx}^2 \end{pmatrix},$$

(16)

where $D$ and $\mu$ are the diffusion coefficient and mobility of the carriers, $E$ is the electric field in-plane of the QW, and $B_{ij}$ describe the effects of the spin-orbit interactions. The latter parameters are functions of the spin-orbit coupling coefficients *α* and *β*, and the transport direction with respect to the crystallographic axes [47]. The evolution of the spin polarization vector is characterized by the dissipation (loss of the spin polarization) due to random spatial motion of carriers and superimposed coherent spin precession. The symmetry of the first mechanism is specified by the geometry of the structure, while the latter is determined by the directions of applied fields. The spin dynamics in such a system can be strongly anisotropic [15,53,56]. For example, for a QW grown in the (001) direction, if the spin-orbit coupling constants *α* and *β* are equal, the spin dissipation is suppressed for electrons propagating along the $(1\bar{1}0)$ direction. For an arbitrary orientation of the electron transport, and *α* = *β*, the solution for the spin polarization density in an appropriately selected coordinate system can be represented as [47]

$$P_x(x) = P_x(0) e^{-\left(\frac{\mu E}{2D} + \sqrt{\left(\frac{\mu E}{2D}\right)^2 + B_{yz}^2}\right)x} \cos(B_{xz}x),$$

$$P_y(x) = P_y(0) e^{-\left(\frac{\mu E}{2D} + \sqrt{\left(\frac{\mu E}{2D}\right)^2 + B_{yz}^2}\right)x} \sin(B_{xz}x), \quad (17)$$

$$P_z(x) = P_z(0) e^{-\left(\frac{\mu E}{2D} + \left|\frac{\mu E}{2D}\right|\right)x},$$

where the angular dependence is hidden in the coefficients $B_{ij}$. According to Eq. (17), for the z-component of the spin polarization density the spin relaxation is suppressed. The transverse component of the spin polarization rotates about the effective magnetic field with the spin precession length, $L_p = 2\pi / B_{xz}$, and decays with the spin dephasing length,

$$L_\perp = \left(\frac{\mu E}{2D} + \sqrt{\left(\frac{\mu E}{2D}\right)^2 + B_{yz}^2}\right)^{-1}. \quad (18).$$

In Fig. 2, calculated angular dependences of the spin precession and spin dephasing lengths are shown for a 10 nm AlGaAs/GaAs/AlGaAs QW. The in-plain electric field in this case affects the spin dephasing mechanism only. Within the utilized model the spin dephasing is suppressed if the applied electric field is along the ($1\bar{1}0$) direction. The frequency of the coherent spin precession is maximal in this case.

**IV. Kinetic transport equations**

Similarly to charge transport, it is possible to describe transport of spin polarization using Boltzmann-like kinetic equations. This can be done within the density matrix approach [57], or methods of non-equilibrium Green's functions [58,59], or Wigner functions [60, 47], where spin property is accounted for starting from quantum mechanical equations. For example, spin-dependent interactions can be described using the Method of Invariants [61]. In this method effects of non-equilibrium spin polarization and/or external magnetic fields naturally appear in the effective mass Hamiltonian as corrections of the same order of magnitude as the effect of conduction band non-parabolicity. The transport equations accounting for spin polarization can be represented as a set of equations for a spin distribution function (for spin ½ it is a set of three equations) plus equation for particle distribution function. The later can be also represented by a set of equations if several types of carriers are utilized. In general case, these equations are coupled.

In this section, we consider a set of kinetic equations for transport of spin-polarized electrons in a semiconductor QW with spin dynamics controlled by the spin-orbit interaction, Eqs. (3)-(4). It can be derived based on the Wigner function approach [47,60]. In the effective mass approximation the single electron Hamiltonian is

$$H = \frac{\mathbf{p}^2}{2m^*} + V(\mathbf{r}) + H_{SO},$$

$$H_{SO} = \mathbf{p} \cdot \vec{\mathfrak{M}} \cdot \boldsymbol{\sigma} / \hbar.$$

(19)

where $\mathbf{p}$ is the electron momentum, $V(\mathbf{r})$, corresponds to the interaction with an electric field oriented in the plane of the QW. The correction due to the spin-orbit interaction, $H_{SO}$, is written in a form, linear in an electron momentum, where $\vec{\mathfrak{M}}$ is a matrix (dyadic) of coupling coefficients [47]. Contribution to the Hamiltonian (19) due to interaction of $\alpha$-component of an electron momentum with $\beta$-component of its spin is proportional to the matrix element $\mathfrak{M}_{\alpha\beta}$. For simplicity, we consider transport of electrons in the one subband approximation with parabolic energy dispersion. The Wigner function for a single electron with spin is [60]

$$W_{s's}(\mathbf{R},\mathbf{k},t) = \int \psi^*(\mathbf{R}-\Delta\mathbf{r}/2,s')\psi(\mathbf{R}+\Delta\mathbf{r}/2,s)e^{-i\mathbf{k}\Delta\mathbf{r}}d^2\Delta r, \quad (20)$$

where $\psi(r,s)$ is an electron wave function. In the spin space, the Wigner function (20) is a matrix 2 by 2. It can be projected to the set of Pauli matrixes, $\sigma_\alpha$, and the unit matrix, $I$, as [62]

$$W = \frac{1}{2}\left(W_n I + W_{\sigma_\alpha}\sigma_\alpha\right), \quad (21)$$

where $W_n$ corresponds to the non-polarized and $W_{\sigma_\alpha}$ corresponds to the $\alpha$-component of the spin-polarized Wigner function. Following the standard procedure of transformation of the Schrödinger equation with the Hamiltonian (19) to the equation for the Wigner function [63], and assuming that the potential, $V(\mathbf{r})$, varies slowly and smoothly with the position $\mathbf{r}$, one can obtain the transport equation for a single electron with spin

$$\frac{\partial W}{\partial t} + \frac{1}{2}\left\{v_j, \frac{\partial W}{\partial x_j}\right\} - \frac{1}{\hbar}\frac{\partial V}{\partial x_j}\frac{\partial W}{\partial k_j} + ik_j\left[v_j, W\right] = \text{St}W, \quad (22)$$

where $v_j = \partial H / \partial p_j$ is the velocity operator. At the right hand side of Eq. (22), the phenomenological scattering term, St$W$, is included. The form of this term can be rather complicated and include transitions between different spin states [57,59].

Operations [A,B] and {A,B} denote commutator and anticommutator, respectively [47,60].

Analytical solutions of a set of spin polarized transport equations can be found for very simple cases only [60]. For quasi-equilibrium transport, approximate solutions can be obtained by an iteration procedure [57] or within a moment expansion scheme [47,58]. If the transport regime far from equilibrium or when electron-electron interactions have to be included, numerical solution schemes [59,64] can be utilized.

**V. Monte Carlo approach**

Monte Carlo simulation approach is one of most powerful methods to study characteristics of transport beyond quasi-equilibrium approximations such as drift-diffusion or linear response approximations. This method has been widely used for modeling of charge carrier transport in semiconductor structures and modern devices [65-67]. Due to its flexibility the approach can easily accommodate in different combinations details of scattering mechanisms, specific device design, material properties and boundary conditions in the simulation (for details see Ref. [67]). However, models designed for quantitative evaluations of transport parameters can be rather time consuming and require a lot of computational resources. Therefore, in many applications where qualitative description of spin-dependent phenomena is required, simplified simulation schemes are useful.

The conventional ensemble Monte Carlo scheme utilized for electronic device design describes transport of classical "representative" particles. Usually, each simulated particle represents a group of real electrons or holes with similar characteristics. In simulation, between scattering events each particle propagates along a classical "localized" trajectory and is affected by external fields. The electric interaction between charge carriers can be accounted for in the mean field approximation. In this case the electric field generated by the non-uniform charge distribution is recalculated every sampling time step using the Poisson equation. The simulation is carried out in a step-wise procedure that consists of "free flight" of a particle in constant external fields according to the classical equations of motion and instantaneous update of external fields at sampling events or update of the energy and momentum of a particle at scattering events. The scattering events are determined by defects, phonons, device geometry, etc., with corresponding scattering rates given by the Fermi golden rule.

The spin property can be incorporated into this scheme easily as an additional parameter, spin polarization vector [68] or spin density matrix [69] calculated for each particle. If spin-dependent interactions between the carriers (dipole-dipole interaction, exchange interaction) are small, then each spin can be considered separately driven by external fields or affected by spin-dependent scattering. Therefore, spin dynamics of each particle can be simulated within the stepwise scheme described in the previous paragraph. In the density matrix representation, during the free flight spin density matrix of the $i^{th}$ particle evolves coherently according to

$$\rho_i(t+\delta t) = e^{-\frac{iH_S\delta t}{\hbar}} \rho_i(t) e^{\frac{iH_S\delta t}{\hbar}} , \quad (23)$$

where $H_S$ is the (spin-dependent) Hamiltonian, assumed *constant* for short time steps, and it changes instantaneously,

$$\rho_i(t) \rightarrow \rho'_i(t) , \quad (24)$$

at spin scattering events.

As characteristic parameters one can use spin the polarization density,

$$P_\alpha = \sum_i Tr(\sigma_\alpha \rho_i) , \quad (25)$$

and spin current densities [70]

$$J_\alpha^\beta = \sum_i v_\beta^i Tr(\sigma_\alpha \rho_i) , \quad (26)$$

where $v_\beta^i$ is the $\beta$ velocity component of $i^{th}$ particle, and sums are taken over all particles located in the grid element of volume, $dV$, at the position **r**. These quantities, Eqs. (25)-(26), can be also given in a normalized form [70] as

$$P_\alpha = \sum_i Tr(\sigma_\alpha \rho_i) \Big/ \sum_i Tr(\rho_i) , \quad X_\alpha^\beta = \sum_i v_\beta^i Tr(\sigma_\alpha \rho_i) \Big/ \sum_i v_\beta^i Tr(\rho_i) . \quad (27)$$

With some variations the described scheme has been successfully utilized to study different properties of spin transport in semiconductors during the past several years [68-83]. Most of the investigations were devoted to properties of electron spin-polarized transport in semiconductor heterostructures where spin dynamics is driven by the spin-orbit interaction [68-77,79,81-83]. Among different aspects addressed in these studies were the effect of an electric gate on the coherent precession of spin polarization in a semiconductor QW [68] and influence of device geometry (effects of width [71,72], length [69], nano-patterns [73], crystallographic symmetry [53], contact/device interface [70,82]) on spin evolution. Investigation of spin transport properties in semiconductor nanowires has been carried out in [74-76] where the authors studied spin dephasing [74,75] and spin noise [76] in a high field transport regime using a multiple-subband model. Details of the precessional spin dephasing were investigated in [50,77,78,83]. An approach to suppressing the effect of spin dephasing was proposed in [79], where it was shown that the electron spin relaxation time depends on the initial spin polarization profile and a specific structure, a "spin-coherence-standing wave," has a several times longer spin relaxation time than the homogeneous electron spin polarization.

In the referenced works, the spin dynamics was described by coherent evolution of spin during free flight of carriers according to Eq. (23) while scattering events were assumed spin independent. An example of Monte Carlo simulation of spin-polarized transport where spin dynamics is controlled by spin scattering events can be found in [80]. In all the discussed studies the back reaction of the spin motion on the particle spatial motion was neglected.

*Relaxation of electron spin polarization near the edge of 2DEG.*

Accordingly to the above discussion, spin transport modeling can be, generally, separated into two parts: simulation of electron spatial motion (usually, neglecting the spin degree of freedom) and calculation of the corresponding spin dynamics. The electron space motion can be considered at different levels of complexity. Clearly the most detailed approach gives the most exact predictions and precise results. However, in most applications high precision is not as important as the speed of the calculations. In this subsection we consider an example based on a simplified consideration of the electron spatial motion and density-matrix approach to the spin. This method was originally proposed in [72]. It was subsequently used for studies of spin relaxation in 2DEG with an antidot lattice [73], dynamics of spin relaxation near the edge of the 2DEG [50], control of spin polarization by pulsed magnetic fields [83], and investigation of long-lived spin coherence states [79].

Within the Monte Carlo simulation algorithm, the space motion of the 2DEG electrons is considered to be along classical (linear) trajectories interrupted by the bulk scattering events. The modelling involves spin-independent bulk scattering processes, which could be caused, e.g., by phonon scatterings or impurities. For the sake of simplicity, the scattering due to such events is assumed to be elastic and isotropic, i.e., the magnitude of the electron velocity is conserved in the scattering events, while the final direction of the velocity vector is randomly selected. The time scale of the bulk scattering events can then be fully characterized by a single rate parameter, the momentum relaxation time, $\tau_p$. It is connected to the mean free path by $L_p = v\tau_p$. Here $v$ is the mean electron velocity. The dynamics of the spin degree of freedom, described by the polarization vector, **P**, in the presence of spin-orbit interactions, reduces to precessions about an effective momentum-dependent magnetic field.

At the initial moment of time the electron coordinates and direction of velocity are randomly generated, while the spin direction is selected according to initial conditions. The main loop of the Monte Carlo simulation algorithm involves the following steps: generation of a time interval between two consecutive scattering events, calculation of the spin dynamics (using the spin polarization vector equation of motion), and random generation of a new direction of the electron velocity after scattering.

D'yakonov-Perel' (DP) spin relaxation mechanism [84] is the leading spin relaxation mechanism in many important experimental situations. In the framework of the DP theory, the initially electron spin polarization relaxes with time. The DP theory was formulated for the bulk of a sample. Considering electron spin relaxation near the edge of 2DEG, one would expect a similar relaxation scenario. This expectation, however, is not correct. Recent studies [50] have demonstrated that the spin relaxation dynamics near the edge is rather unusual and can not be described by a simple exponential law, as follows from the DP theory. The described Monte Carlo simulation algorithm was used complementary to analytical investigation of short-time spin relaxation dynamics [50]. Fig. 3 shows the electron spin polarization at different moments of time. Whereas the electron spin polarization exponentially decreases in the bulk region, its behavior near the edge is rather unusual. Longer spin relaxation time near the edge, spin polarization oscillations and spin polarization transfer from the perpendicular to in-plane component were observed.

*Spin-polarized transport in a finite length device structure.*

The size of modern semiconductor devices is submicron [1]. In such structures the average electric field can easily reach several kV/cm, and transport of charge carriers is characterized by non-equilibrium charge distribution, potential profile, electron energy, etc. To study spin dynamics in such a regime, transport of spin-polarized electrons in an asymmetric quantum well formed in a $In_{0.52}Al_{0.48}As/In_{0.53}Ga_{0.47}As/In_{0.52}Al_{0.48}As$ semiconductor heterostructure, Fig. 4, has been simulated using an ensemble Monte Carlo scheme [81].-

The length of the device was taken $l = 0.55\,\mu m$. For study of spin transport, the width was assumed infinite, neglecting effects of a finite device width studied in [71,72]. The boundary conditions were specified by the following rules. Thermalized electrons were generated at the left (source) boundary. The charge neutrality of the whole device was conserved during the simulation. Once an electron left the device, a new particle was introduced at the source boundary. Initially, injected particles were assumed 100% spin polarized in a specific direction of polarization. The spin polarization of a single electron was defined by the single electron spin density matrix, Eq. (12). The particle transport was simulated consistently with using the Poisson equation to update the charge distribution along the device channel. The spin dynamics was evaluated for each single particle using Eq. (23), where spin dependent Hamiltonian, $H_S$, was represented by two spin-orbit terms, Eqs. (3)-(4). In this case, Eq. (23) involves

$$e^{-iH_S \delta t/\hbar} = \begin{pmatrix} \cos(|\gamma|\delta t) & i\frac{\gamma}{|\gamma|}\sin(|\gamma|\delta t) \\ i\frac{\gamma^*}{|\gamma|}\sin(|\gamma|\delta t) & \cos(|\gamma|\delta t) \end{pmatrix}, \qquad (28)$$

and its Hermitian conjugate. The parameter $\gamma$ is determined by spin-orbit coupling constants and an electron momentum as

$$\gamma = \hbar^{-1}\left[\left(\alpha k_y + \beta k_x\right) + i\left(\alpha k_x + \beta k_y\right)\right] \ . \tag{29}$$

In the simulation, the total number of representative particles was $N = 55000$, and the sampling time step was $\Delta t_{samp} = 1$ fsec. The program was run for 20000 time steps to achieve the steady-state transport regime, and then data were collected during the last 2000 time steps.

As shown in Fig. 5, the electrons injected with the average thermal energy 26 meV are rapidly heated up if the voltage $V_{DS} = 0.1$-$0.25$ V is applied along the device channel. The electron concentration is nearly constant in the channel except of a small accumulation region (~ 0.01 μm) that is defined by the boundary conditions. Fig. 6 illustrates that the spin dephasing in the structure is somewhat dependent on the applied voltage and temperature. Similar features have been emphasized in [74] where spin transport in GaAs nanowires was studied.

*Spin injection through a Schottky barrier.*

The problem of spin injection into a non-magnetic semiconductor is one of important issues of semiconductor spintronics. Design of most of the proposed spintronic devices [13,15-17,19] requires an efficient source of spin-polarized carriers. Recently, it has been proven theoretically [85-87] and demonstrated experimentally [88-90] that spin-polarized electrons can be injected into a semiconductor from a metal ferromagnetic contact through a Schottky barrier. The Monte Carlo approach has been utilized to investigate spin dynamics in a device that consists of a ferromagnetic metal contact connected to a semiconductor heterostructure similar to that is shown in Fig. 4 [70,82]. In the metal part the electrons are assumed spin-polarized with some distribution of polarization as a function of the electron energy, while in the semiconductor part of the device spin dynamics is controlled by the spin-orbit interaction. The conduction band profile of the considered device, calculated based on the electron concentration in a semiconductor with boundary conditions specified by the applied voltage is shown in Fig. 7. Because of the potential barrier at the metal/semiconductor interface, the boundary conditions, corresponding to this design, are appreciably different from those utilized in the previous example. Details of the model can be found in Ref. [70].

In the simulation, spin-polarized electrons are transported into the semiconductor by two different mechanisms, thermionic emission and tunneling. Injected electrons represent only a small fraction of electrons in the semiconductor part. Therefore, two types of representative particles, injected and persistently existing in the device channel, have been used in the model. In Fig. 8 the energy distribution of the injected electrons at bias voltage $V_{DS} = 0.1$ V at room temperature is shown.

Because of high kinetic energy of electrons injected [70], the spin-orbit interaction was accounted for beyond the linear approximation, Eqs. (3)-(4).

The particles persistently existing in the semiconductor part do not affect charge and spin current densities. However, they produce a non-polarized background that suppresses the spin polarization density, Eq. (25), in the channel. Therefore, the spin current density, Eq. (4), was considered as a more useful characteristic of the spin dynamics in the structure. In Fig. 9, the spin current density is shown for two different orientations of the injected polarization. Because of spin dephasing, spin current is not conserved along the channel. The spin dynamics is strongly anisotropic with respect to the injected polarization. It has been found that spin current polarized in the plane of the QW, orthogonal to the transport direction, is conserved for a longer distance [70]. It has also been shown that by $n^+$ doping of the Schottky barrier region spin current can be controlled over several orders of magnitude [82].

**Acknowledgments**

We gratefully acknowledge helpful discussions and collaboration with M.-C. Cheng, J. A. Nesteroff, E. Shafir, M. Shen and I. D. Vagner. This research was supported by the National Security Agency and Advanced Research and Development Activity under Army Research Office contract DAAD-19-02-1-0035, and by the National Science Foundation, grant DMR-0121146.

# References

1. International Technology Roadmap for Semiconductors (Ed. 2003), website http://public.itrs.net/Files/2003ITRS/Home2003.htm

2. Research needs for novel devices (SRC Edition May 2003)

3. Prinz, G.: 'Spin-Polarized Transport', *Phys. Today*, 1995, **48** (4), pp. 58-63

4. Wolf, S.A., Awschalom, D.D., Buhrman, R.A., Daughton, J.M., von Molnar, S., Roukes, M.L., Chtchelkanova, A.Y., Treger, D.M.: 'Spintronics: a spin-based electronics. Vision for the future', *Science*, 2001, **294**, pp. 1488-1495

5. Das Sarma, S., 'Spintronics', *Am. Sci.*, 2001, **89**, pp. 516-523

6. Awschalom, D.D., Flatte M.E., and Samarth, N.: 'Spintronics', *Sci. Am.*, 2002, **286**, pp. 66-73

7. Akinaga, H., and Ohno, H.: 'Semiconductor spintronics', *IEEE Trans. Nanotechnology*, 2002, **1**, pp. 19-31

8. Jonker, B.T.: 'Progress toward electrical injection of spin-polarized electrons into semiconductors', *Proceedings of the IEEE*, 2003, **91**, pp. 727-740

9. Zutic, I., Fabian, J., Das Sarma, S.: 'Spintronics: Fundamentals and applications', *Rev. Mod. Phys.*, 2004, **76**, pp. 323-410

10. Parkin, S., Jiang, X., Kaiser, C., Panchula, A., Roche, K., Samant, M.: 'Magnetically engineered spintronic sensors and memory', *Proceedings of the IEEE*, 2003, **91**, pp. 661-680

11. Fabian, J., Zutic, I., Das Sarma, S.: 'Magnetic bipolar transistor', *Appl. Phys. Lett.*, 2004, **84**, pp. 85-87

12. Flatte, M.E., Yu Z.G., Johnson-Halperin, E., Awschalom, D.D.: 'Theory of semiconductor magnetic bipolar transistors', *Appl. Phys. Lett.*, 2003, **82**, pp. 4740-4742

13. Datta S., Das B.: 'Electronic analog of the electro-optic modulator', *Appl. Phys. Lett.*, 1990, **56**, pp. 665-667

14. Wang, B., Wang, J., and Guo, H.: 'Quantum spin field effect transistor', *Phys. Rev. B*, 2003, **67**, article 092408, pp. 1-4

15. Schliemann, J., Egues, J.C., Loss, D.: 'Nonballistic spin-field-effect transistor', *Phys. Rev. Lett.*, 2003, **90**, article 146801, pp. 1-4

16. Egues, J.C., Burkard, G., Loss, D.: 'Datta-Das transistor with enhanced spin control', *Appl. Phys. Lett.*, 2003, **82**, pp. 2658-2660

17. Wang X.F., and Vasilopoulos P.: 'Influence of subband mixing due to spin-orbit interaction on the transmission through periodically modulated waveguides', *Phys. Rev. B*, 2003, **68**, article 035305, pp. 1-8

18. Hall, K.C., Lau, W.H., Gundogdu, K., Flatte, M.E., Boggess, T.F.: 'Non-magnetic semiconductor spin transistor', *Appl. Phys. Lett.*, 2003, **83**, pp. 2937-2939

19. Jonker, B.T.: 'Polarized optical emission due to decay or recombination of spin-polarized injected carriers', Feb. 23, 1999, US patent 5874749

20. Mani, R.G., Johnson, W.B., Narayanamurti, V., Privman, V., and Zhang, Y.-H., 'Nuclear spin based memory and logic in quantum Hall semiconductor nanostructures for quantum computing applications', *Physica E*, 2002, **12**, pp. 152-156

21. Vrijen, R., Yablonovitch, E., Wang, K., Jiang, H.W., Balandin, A., Roychowdhury, V., Mor, T., DiVincenzo, D.: 'Electron-spin-resonance transistors for quantum computing in silicon-germanium heterostructures', *Phys. Rev. A*, 2000, **62**, article 012306, pp. 1-10

22. Bandyopadhyay, S., Cahay, M.: 'Proposal for a spintronic femto-Tesla magnetic field sensor', *Physica E*, 2005, **27**, pp. 98-103

23. Ciuti, C., McGuire, J.P., Sham, L.J.: **'**Spin-dependent properties of a two-dimensional electron gas with ferromagnetic gates', *Appl. Phys. Lett.*, 2002, **81**, pp. 4781-4783

24. Osipov, V.V., Bratkovsky, A.M.: 'A class of spin injection-precession ultrafast nanodevices', *Appl. Phys. Lett.*, 2004, **84**, pp. 2118-212025. Bratkovsky, A.M., Osipov, V.V.: 'High-Frequency Spin-Valve Effect in a Ferromagnet-Semiconductor-Ferromagnet Structure Based on Precession of the Injected Spins', *Phys. Rev. Lett.*, 2004, **92**, article 098302, pp. 1-4

26. D'yakonov, M., in *Future Trends in Microelectronics: The Nano, the Giga, and the Ultra*, edited by S. Luryi, J. Xu, A. Zaslavsky (Wiley-IEEE Press, 2004)

27. Bandyopadhyay, S., Cahay, M.: 'Reexamination of some spintronic field-effect device concepts', *Appl. Phys. Lett.*, 2004, **85**, pp. 1433-1435

28. Ohno, Y., Terauchi, R., Adachi, T., Matsukura, F., Ohno, H.: 'Spin relaxation in GaAs (110) quantum wells', *Phys. Rev. Lett.*, 1999, **83**, pp. 4196-4199

29. Karimov, O.Z., John, G.H., Harley, R.T., Lau, W.H., Flatté, M.E., Henini, M., Airey, R.: 'High temperature gate control of quantum well spin memory', *Phys. Rev. Lett.*, 2003, **91**, article 246601, pp. 1-4

30. Malajovich, I., Berry, J.J., Samarth, N., Awschalom, D.D.: 'Persistent sourcing of coherent spins for multifunctional semiconductor spintronics', *Nature*, 2001, **411**, pp. 770-772

31. Kikkawa, J.M., Awschalom, D.D.: 'Lateral drag of spin coherence in gallium arsenide', *Nature*, 1999, **397**, pp. 139-141

32. Rudolph, J., Hägele, D., Gibbs, H.M., Khitrova, G., Oestreich, M.: 'Laser threshold reduction in a spintronic device', *Appl. Phys. Lett.*, 2003, **82**, pp. 4516-4518

33. Lannon, J.M.Jr., Dausch, D.E., Temple, D.: 'High sensitivity polarized-light discriminator device', Apr. 24, 2003, US patent application

34. Stevens, M.J., Smirl, A.L., Bhat R.D.R., Najmaie, A., Sipe J.E., van Driel H.M.: 'Quantum interference control of ballistic pure spin currents in semiconductors', *Phys. Rev. Lett.*, 2003, **90**, article 136603, pp. 1-4

35. Landau L.D., and Lifshitz, E.M.: 'Quantum Mechanics' (Butterworth-Heinemann, Oxford, 1997)

36. Condon, E.U., and Shortley, G.H.: 'The Theory of Atomic Spectra' (Cambridge University Press, Cambridge, 1953)

37. Dresselhaus, G.: 'Spin-orbit coupling effects in zinc blende structures', *Phys. Rev.*, 1955, **100**, pp. 580-586

38. Bychkov, Yu., and Rashba, E.I.: 'Oscillatory effects and the magnetic susceptibility of carriers in inversion layers', *J. Phys. C*, 1984, **17**, pp. 6039-6045

39. Gantmakher, V.F., Levinson, Y.B.: 'Carrier scattering in metals and semiconductors' in 'Modern Problems in Condensed Matter Science' v. 19. Series editors Agranovich, V.M., Maradudin, A. A. (North-Holland, New York 1987)

40. Yu, Z.G., Flatté, M.E.: 'Spin diffusion and injection in semiconductor structures: Electric field effects', *Phys. Rev. B*, 2002, **66**, article 235302, pp. 1-14

41. Yu, Z.G., Flatté, M.E.: 'Electric-field dependent spin diffusion and spin injection into semiconductors', *Phys. Rev. B*, 2002, **66**, article 201202, pp. 1-4

42. Žutić, I., Fabian, J., Das Sarma, S.: 'Spin-polarized transport in inhomogeneous magnetic semiconductors: theory of magnetic/nonmagnetic *p-n* junctions', *Phys. Rev. Lett.*, 2002, **88**, article 066603, pp. 1-4

43. Pershin, Yu.V., Privman, V.: 'Focusing of spin polarization in semiconductors by inhomogeneous doping', *Phys. Rev. Lett.*, 2003, **90**, article 256602, pp. 1-4

44. Pershin, Yu.V., Privman, V.: 'Propagation of spin-polarized electrons through interfaces separating differently doped semiconductor regions', *Proc. Conference "IEEE-NANO 2003"* (IEEE Press, Monterey, CA, 2003), pp. 168-170

45. Martin, I.: 'Spin-drift transport and its applications', *Phys. Rev. B*, 2003, **67**, article 014421, pp. 1-5

46. Pershin, Yu.V.: 'Accumulation of electron spin polarization at semiconductor interfaces', *Phys. Rev. B*, 2003, **68**, article 233309, pp. 1-4

47. Saikin, S.: 'Drift-diffusion model for spin-polarized transport in a non-degenerate 2DEG controlled by a spin-orbit interaction', *J. Phys.: Condens. Matter.*, 2004, **16**, pp. 5071-5081

48. Culcer, D., Sinova, J., Sinitsyn, N.A., Jungwirth, T., MacDonald, A.H., Niu, Q.: 'Semiclassical Spin Transport in Spin-Orbit-Coupled Bands', *Phys. Rev. Lett.*, 2004, **93**, article 046602, pp. 1-4

49. Pershin, Yu.V.: 'Drift–diffusion approach to spin-polarized transport', *Physica E*, 2004, **23**, pp. 226-231

50. Pershin, Yu.V.: 'Dynamics of spin relaxation near the edge of two-dimensional electron gas', *Physica E*, 2005, **27** pp. 77-81

51. Burkov, A.A., Nunez, A.S., MacDonald, A.H.: 'Theory of spin-charge-coupled transport in a two-dimensional electron gas with Rashba spin-orbit interactions', *Phys. Rev. B*, 2004, **70**, article 155308, pp. 1-8

52. Shafir, E., Shen, M., Saikin, S.: 'Modulation of spin dynamics in a channel of a nonballistic spin-field effect transistor', *Phys. Rev. B*, 2004, **70**, article 241302(R), pp. 1-4

53. Saikin, S., Shen, M., and Cheng, M.-C.: 'Study of spin-polarized transport properties for spin-FET design optimization', *IEEE Trans. Nanotechnology*, 2004, **3**, pp. 173-179

54. Dyakonov, M.I., and Kachorovskii, V.Yu.: 'Spin relaxation of two-dimensional electrons in noncentrosymmertic semiconductors', *Sov. Phys. Semicond.*, 1986, **20**, pp. 110-112

55. Bauer, G. E. W., Tserkovnyak, Y., Huertas-Hernando, D., Brataas, A.: 'From digital to analogue magnetoelectronics: theory of transport in non-collinear magnetic nanostructures,' in 'Advances in Solid State Physics,' vol. 43 (Springer-Verlag, Berlin 2003)

56. Averkiev, N. S., Golub, L. E.: 'Giant spin relaxation anisotropy in zinc-blende heterostructures', *Phys. Rev. B*, 1999, **60**, pp. 15582-15584

57. Ivchenko, E.L., Lyanda-Geller, Yu.B., Pikus, G.E.: 'Current of thermalized spin-oriented photocarriers', *Sov. Phys. JETP*, 1990, **71**, pp. 550-557

58. Takahashi, Y., Shizume, K., Masuhara, N.: 'Spin diffusion in a two-dimensional electron gas', *Phys. Rev. B*, 1999, **60**, pp. 4856-4865

59. Weng, M.Q., Wu, M.W.: 'Kinetic theory of spin transport in *n*-type semiconductor quantum wells', *J. Appl. Phys.*, 2003, **93**, pp. 410-420

60. Mishchenko, E.G., Halperin, B.I.: 'Transport equations for a two-dimensional electron gas with spin-orbit interaction', *Phys. Rev. B*, 2003, **68**, article 045317, pp. 1-6

61. Bir, G.L., Pikus, G. E.: 'Symmetry and strain-induced effects in semiconductor' (Keter Publishing House Jerusalem Ltd. 1974)

62. Carruthers, P., Zachariasen, F.: 'Quantum collision theory with phase-space distributions', *Rev. Mod. Phys.*, 1983, **55**, pp. 245-285

63. Wigner, E.: 'On the quantum correction for thermodynamic equilibrium', *Phys. Rev.*, 1932, **40**, pp. 749-759

64. Weng, M.Q., Wu, M.W., Jiang, L.: 'Hot-electron effect in spin dephasing in *n*-type GaAs quantum wells', *Phys. Rev. B*, 2004, **69**, article 245320, pp. 1-9

65. Hess, K.: 'Monte Carlo Device Simulation: Full Band and Beyond' (Kluwer Academic Publishers, Boston, 1991)

66. Tomizawa, K.: 'Numerical Simulation of Submicron Semiconductor Devices' (Artech House, London, Boston, 1993)

67. Fischetti, M.V., Laux, S.E.: 'DAMOCLES Theoretical Manual' (IBM, Yorktown Heights, 1995)

68. Bournel, A., Dollfus, P., Bruno, P., Hesto, P.: 'Gate-induced spin precession in an $In_{0.53}Ga_{0.47}As$ two dimensional electron gas', *Eur. Phys. J. AP.*, 1998, **4**, pp. 1-4

69. Saikin, S., Shen, M., Cheng, M.-C., and Privman, V.: 'Semiclassical Monte Carlo model for in-plane transport of spin-polarized electrons in III–V heterostructures', *J. Appl. Phys.*, 2003, **94**, pp. 1769-1775

70. Shen, M., Saikin, S., Cheng, M.-C.: 'Monte Carlo modeling of spin injection through a Schottky barrier and spin transport in a semiconductor quantum well', *J. Appl. Phys.*, 2004, **96**, pp. 4319-4325

71. Bournel, A., Dollfus, P., Bruno, P., Hesto, P.: 'Spin-dependent transport phenomena in a HEMT', *Physica B*, 1999, **272**, pp. 331-334

72. Kiselev, A.A., Kim, K.W.: 'Progressive suppression of spin relaxation in two-dimensional channels of finite width', *Phys. Rev. B*, 2000, **61**, pp. 13115-13120

73. Pershin, Yu.V. and Privman, V.: 'Slow spin relaxation in two-dimensional electron systems with antidots', Phys. Rev. B, 2004, **69**, article 073310, pp.1-4

74. Pramanik, S., Bandyopadhyay, S., Cahay, M.: 'Spin dephasing in quantum wires', *Phys. Rev. B*, 2003, **68**, article 075313, pp.1-10

75. Pramanik, S., Bandyopadhyay, S., Cahay, M.: 'Decay of spin-polarized hot carrier current in a quasi-one-dimensional spin-valve structure', *Appl. Phys. Lett.*, 2004, **84**, pp. 266-268 ().

76. Pramanik, S., Bandyopadhyay, S.: 'Spin fluctuations and "spin noise"', *e-print* cond-mat/0312099 (2003)

77. Bournel, A., Dollfus, P., Cassan, E., Hesto P.: 'Monte Carlo study of spin relaxation in AlGaAs/GaAs quantum wells', *Appl. Phys. Lett.*, 2000, **77**, pp. 2346-2348

78. Barry, E.A., Kiselev, A.A., Kim, K.W.: 'Electron spin relaxation under drift in GaAs', *Appl. Phys. Lett.*, 2003, **82**, pp. 3686-3688

79. Pershin, Yu.V.: 'Long-lived spin coherence states', *Phys. Rev. B*, 2005, (in press); *e-print* cond-mat/0311223

80. Pershin, Yu.V., and Privman, V.: 'Spin relaxation of conduction electrons in semiconductors due to interaction with nuclear spins', *Nano Lett.*, 2003, **3**, pp. 695-700

81. Shen, M., Saikin, S., Cheng, M.-C., and Privman, V.: 'Monte Carlo modeling of spin FETs controlled by spin-orbit interaction', *Math. Comp. Simul.*, 2004, **65**, pp. 351-363

82. Shen, M., Saikin, S., Cheng, M.-C.: 'Spin injection in spin FETs using a step-doping profile', *IEEE Trans. Nanotechnology*, 2005, **4**, pp. 40-44.

83. Pershin, Yu. V.: 'Spin coherence control by pulsed magnetic fields', *e-print* cond-mat/0310225

84. D'yakonov, M. I., Perel', V. I.: 'Spin orientation of electronics associated with the interband absorption of light in semiconductors,' *Soviet Phys. JETP*, 1971, **33**, pp. 1053-1059.

85. Rashba, E.I.: 'Theory of electrical spin injection: Tunnel contacts as a solution of the conductivity mismatch problem', *Phys. Rev. B*, 2000, **62**, pp. R16267-R16270

86. Albrecht, J. D., Smith, D.L.: 'Spin-polarized electron transport at ferromagnet/semiconductor Schottky contacts', *Phys. Rev. B*, 2003, **68**, article 035340, pp.1-14

87. Osipov, V. V., Bratkovsky, A. M.: 'Efficient nonlinear room-temperature spin injection from ferromagnets into semiconductors through a modified Schottky barrier', *Phys. Rev. B*, 2004, **70**, article 205312, pp 1-6

88. Ploog, K. H.: 'Spin injection in ferromagnet-semiconductor heterostructures at room temperature', *J. Appl. Phys.*, 2002, **91**, pp. 7256-7260

89. Hanbicki, A.T., van't Erve, O. M. J., Magno, R., Kioseoglou, G., Li, C.H., Jonker, B.T., Itskos, G., Mallory, R., Yasar, M., Petrou, A.: 'Analysis of the transport process providing spin injection through an Fe/AlGaAs Schottky barrier', *App. Phys. Lett.*, 2003, **82**, pp. 4092-4094

90. Adelmann, C., Lou, X., Strand, J., Palmstrøm, C. J., Crowell, P. A.: 'Spin injection and relaxation in ferromagnet-semiconductor heterostructures', *Phys. Rev. B*, 2005, **71**, article 121301(R), pp 1-4

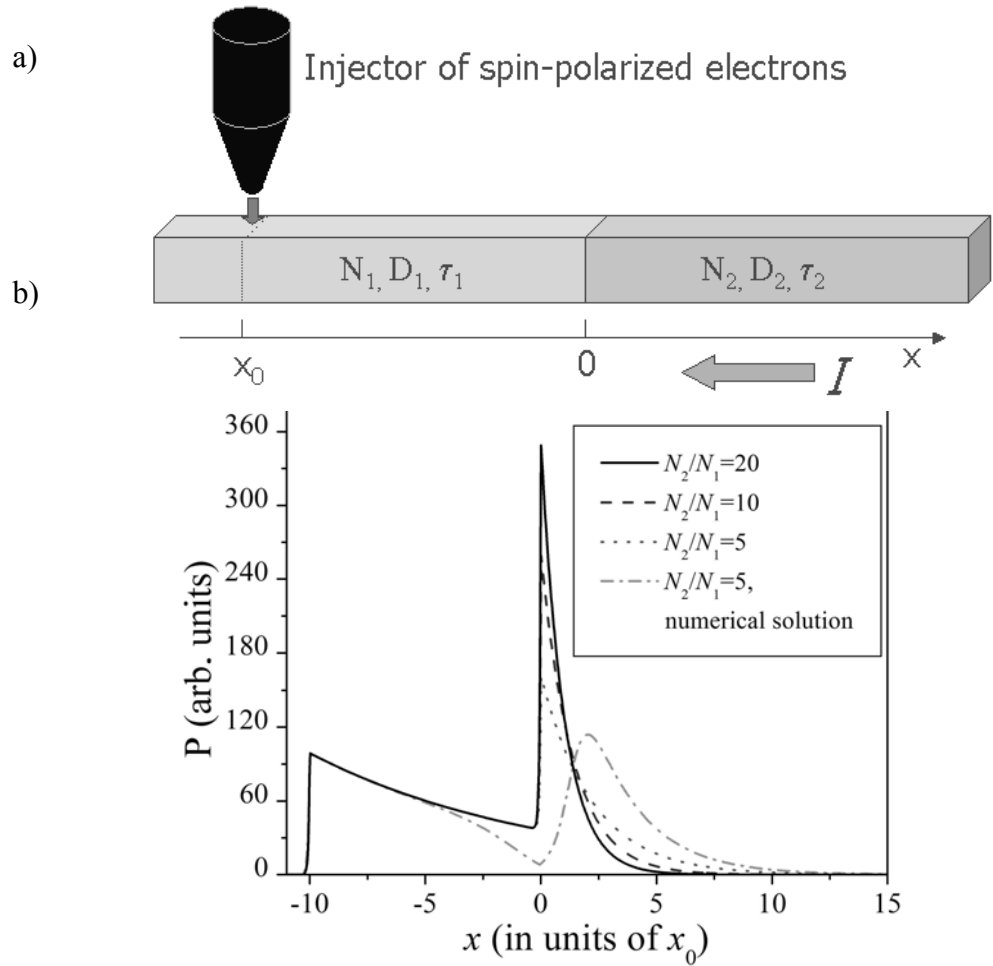

**Figure 1.** Propagation of spin-polarized electrons through a boundary between two n-type semiconductors: (a) schematics of the system: spin-polarized electrons are injected at $x = x_0$ and move toward the interface located at $x = 0$, under the action of the electric field; (b) spin polarization density created by a point source at $x_0 = -10$, as a function of $x$ for different doping densities $N_2$, with $\tau_1 = \tau_2$ and $D_1 = D_2$, showing the increase of the spin accumulation with higher $N_2$. The dash-dotted line represents the spin polarization density obtained with taking into account additionally the redistribution of the electron density near the boundary, as detailed in [43].

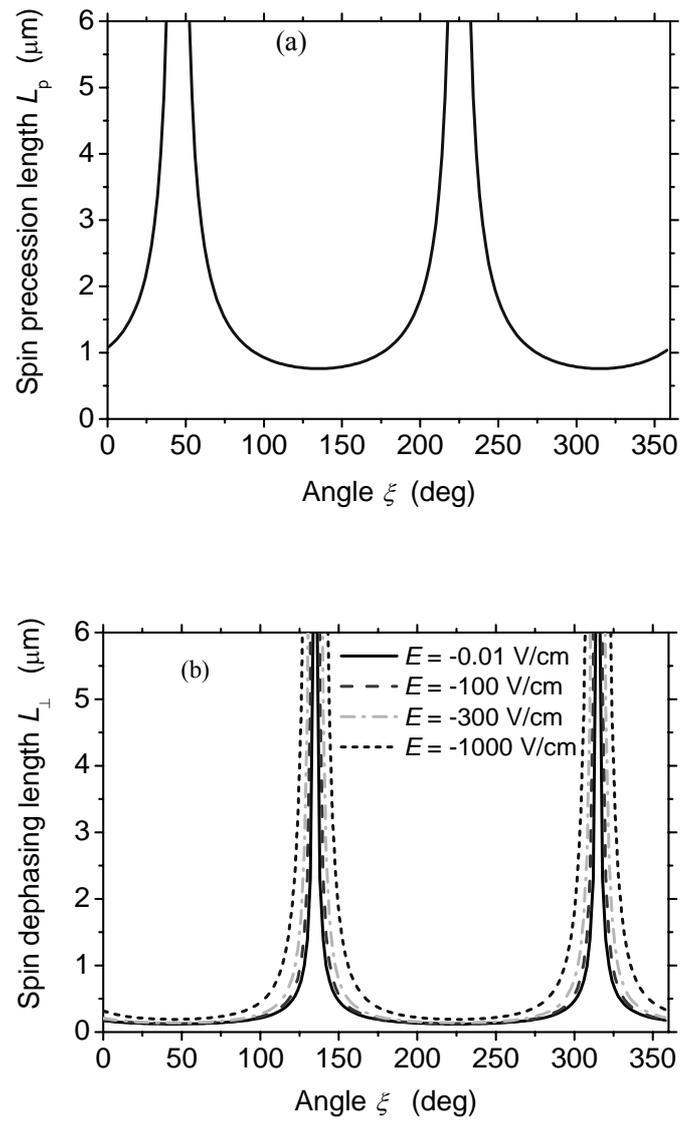

**Figure** 2. Spin precession length (a), and transverse spin dephasing length (b), for different transport orientations with respect to the (001) crystallographic direction, at room temperature.

a)

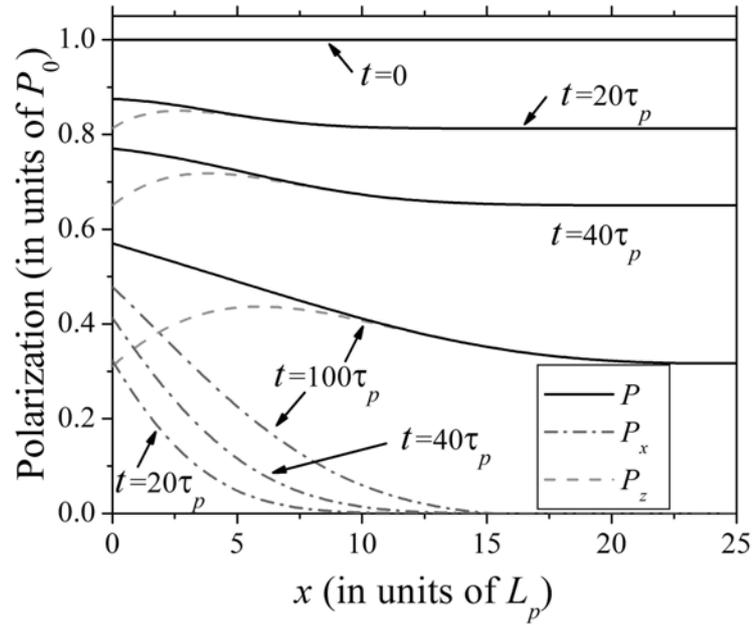

b)

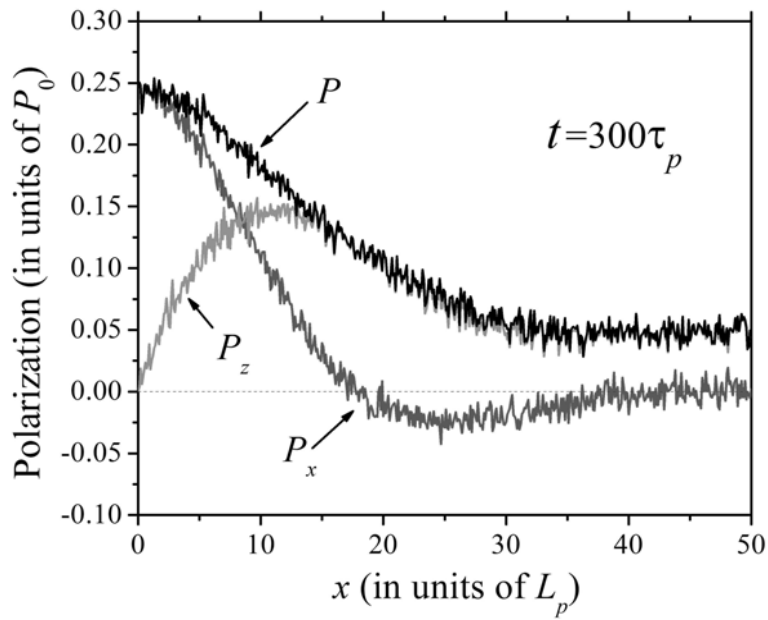

**Figure 3.** (a) Short- and (b) long-time evolution of electron spin polarization near the edge of 2DEG. Suppression of spin relaxation near the edge, as well as spin oscillations in time and space were found.

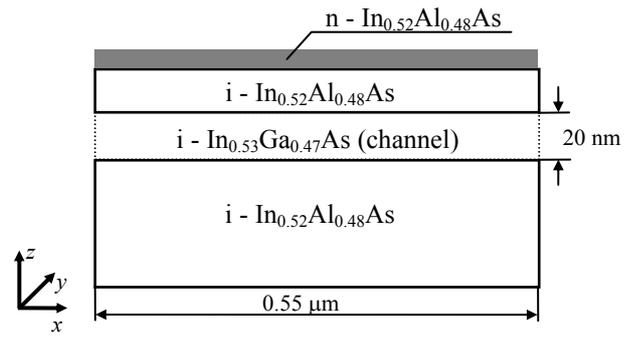

**Figure 4.** Schematics of the device simulated.

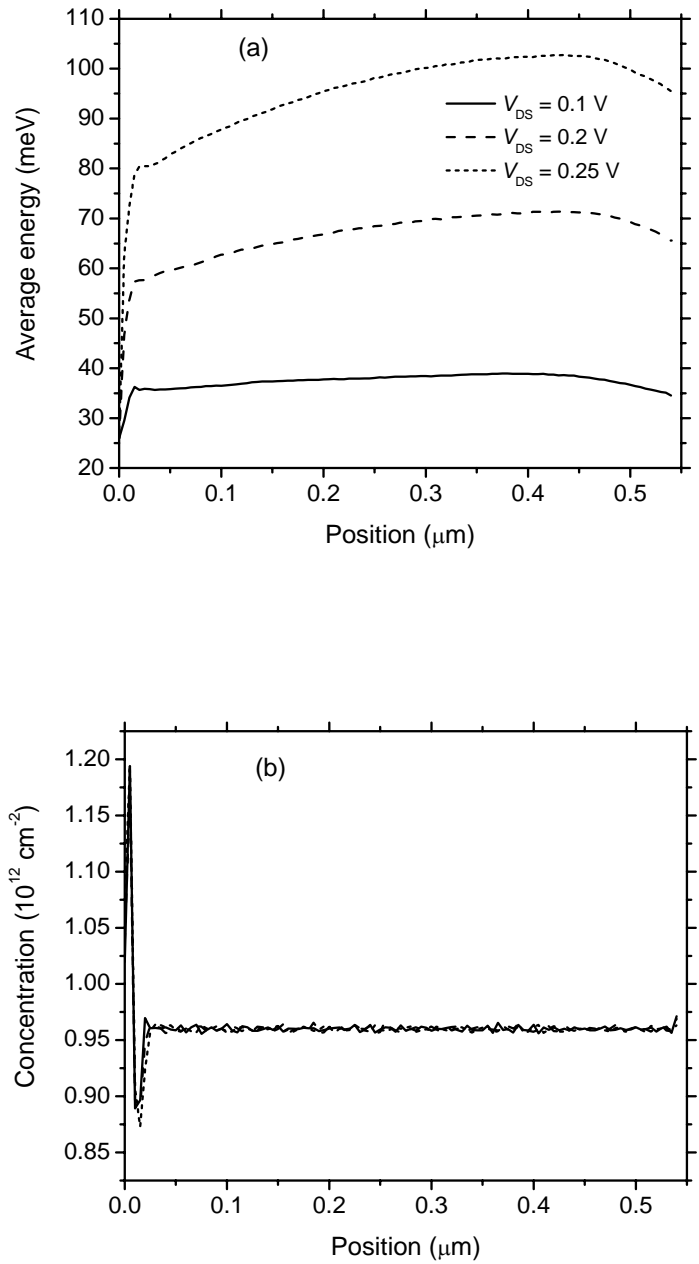

**Figure 5.** Energy profile (a) and electron concentration (b) along the device channel for different values of drain/source voltage, for $T = 300$ K.

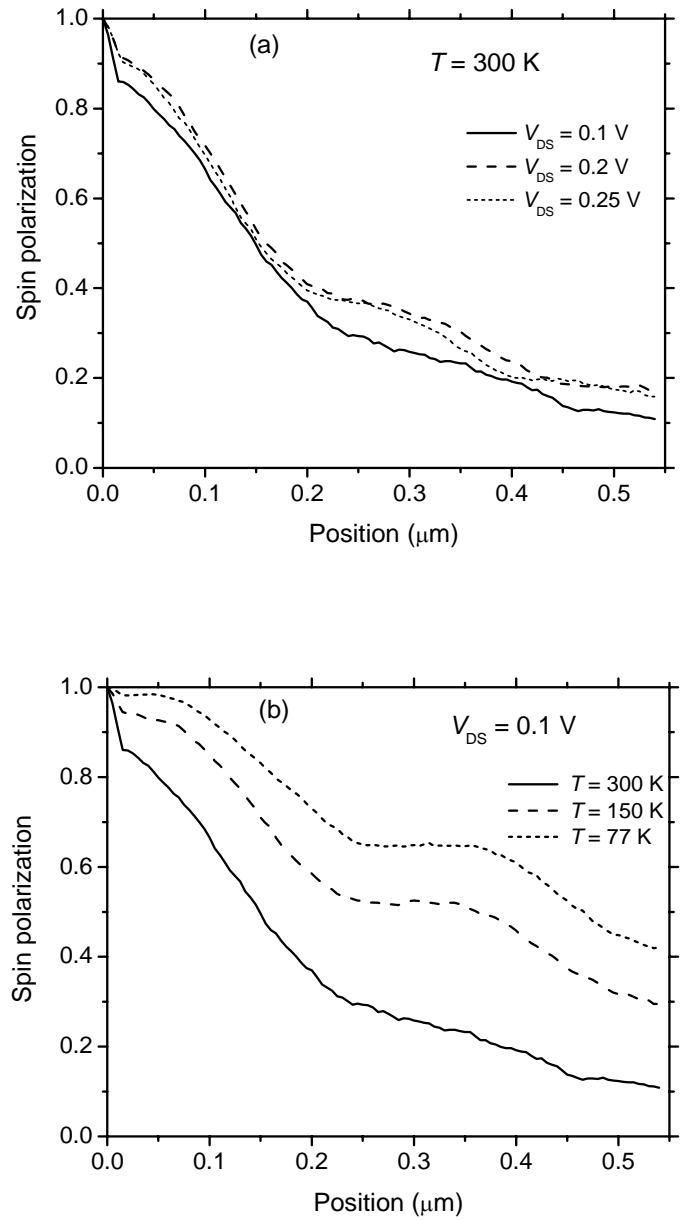

**Figure 6.** (a) Distribution of spin polarization for different values of drain/source voltage, at $T = 300$ K. (b) Distribution of spin polarization for different temperatures for $V_{DS} = 0.1$ V.

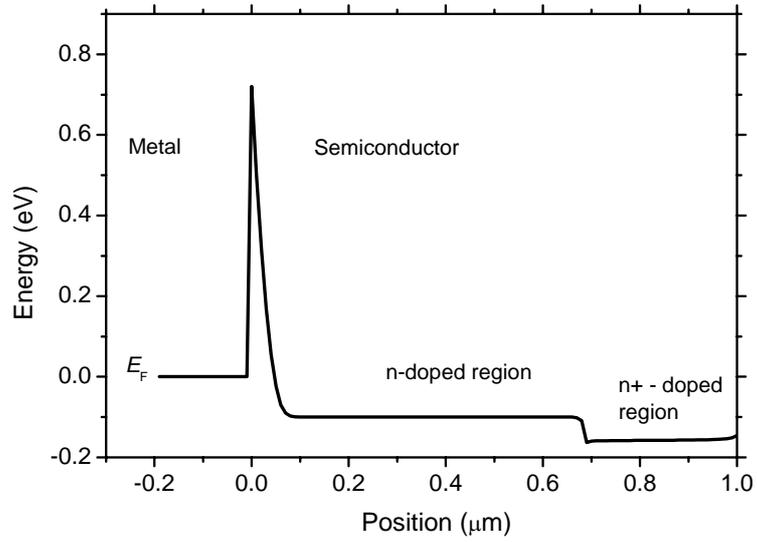

**Figure 7.** Simulated conduction band profile, $V_{DS} = 0.1$ V.

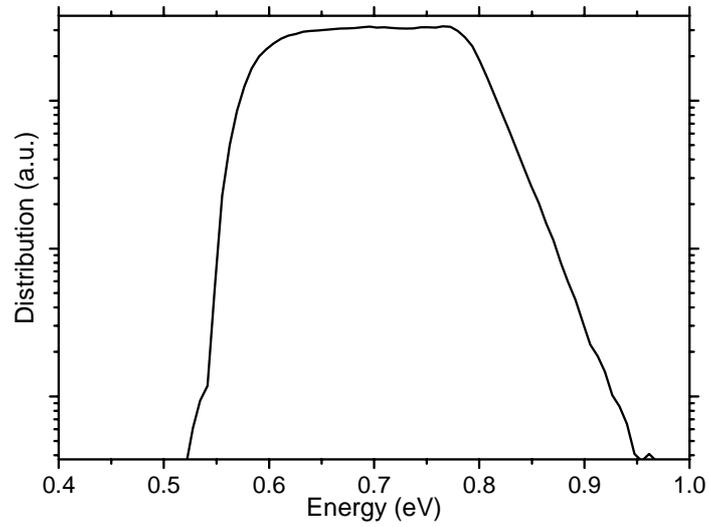

**Figure 8.** Energy distribution of electrons injected through the Schottky barrier, $V_{DS} = 0.1$ V, $T = 300$ K.

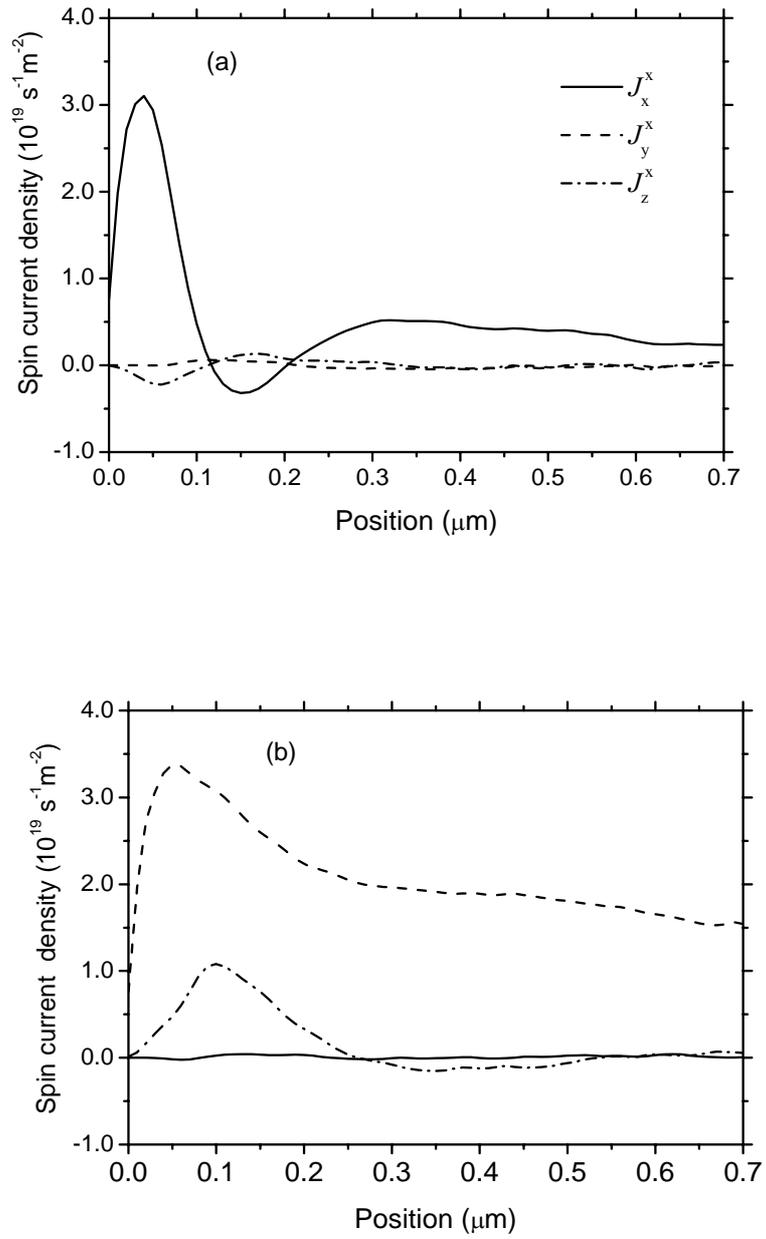

**Figure 9.** Spin current density for two orientations of spin injection, (a) injected polarization is along the channel, (b) injected polarization is in the plane of the QW orthogonal to the channel.